\documentclass[pre,superscriptaddress,twocolumn,showpacs]{revtex4}

\usepackage{graphicx}
\usepackage{amssymb}

\def\ssquare{\mbox{\footnotesize $\square$}}

\def\lcirc{\mbox{\large $\circ$}}

\def\3{\ss }           

\def\bxi{\mbox{\boldmath $\xi$}}

\begin{document}

\title{
Resummed Green-Kubo relations for a fluctuating fluid-particle model} 
\author{T. Ihle}
\affiliation{ Institut f\"ur Computeranwendungen, Universit\"at 
Stuttgart Pfaffenwaldring 27, 70569 Stuttgart, Germany}
\author{E. T{\"u}zel}
\affiliation{School of Physics and Astronomy, 116 Church Street SE, University of Minnesota,  \\ Minneapolis, MN 55455, USA }
\affiliation{Supercomputing Institute, University of Minnesota, 
599 Walter Library, 117 Pleasant Street S.E., \\
Minneapolis, MN 55455, USA}
\author{D.M. Kroll}
\affiliation{Supercomputing Institute, University of Minnesota, 
599 Walter Library, 117 Pleasant Street S.E., \\
Minneapolis, MN 55455, USA}
\date{\today}

\begin{abstract}

A recently introduced stochastic model for fluid flow can be made Galilean 
invariant by introducing a random shift of the computational grid before 
collisions. This grid shifting procedure accelerates momentum transfer between 
cells and leads to a collisional contribution to transport coefficients. 
By resumming the Green-Kubo relations derived in a previous paper, it is 
shown that this collisional contribution to the transport coefficients 
can be determined exactly. The resummed Green-Kubo relations 
also show that there are no mixed kinetic-collisional contributions 
to the transport coefficients. The leading correlation corrections to 
the transport coefficients are discussed, and   
explicit expressions for the transport coefficients are presented and 
compared with simulation data. 

\end{abstract}
\maketitle

A recently introduced stochastic model for fluid flow \cite{male_99,male_00a}
with efficient multi-particle interactions---which we will 
call Stochastic Rotation Dynamics (SRD)---is a promising tool for 
the coarse-grained modeling of a fluctuating solvent, particularly for 
colloidal \cite{inou_02} and polymer suspensions 
\cite{male_00b,falc_03b,ripo_04}. 
SRD can be thought of as a ``hydrodynamic 
heat bath'', the details of which are not fully resolved, but which
provides the correct hydrodynamic interactions between embedded particles.
In addition to its numerical advantages, its simplicity makes 
it possible 
to obtain analytic expressions for the transport coefficients which are 
valid for both large and small mean free paths, something which is 
very difficult to do for other mesoscale particle-based methods.

In its original form \cite{male_99,male_00a}, the SRD algorithm was not 
Galilean invariant at low temperatures, where the mean free path, $\lambda$, 
is smaller than the cell size $a$. However, as was shown in 
Refs. \cite{ihle_01} and \cite{ihle_03a},  
Galilean invariance can be restored by introducing a random shift 
of the computational grid before every multi-particle interaction. 
A discrete-time projection operator technique was then used \cite{ihle_03a} 
to derive the Green-Kubo (GK) relations for the model's transport coefficients. 
Using these results, explicit expressions for the transport coefficients were 
derived in an accompanying paper \cite{ihle_03b}. In particular, it was shown 
that the grid shifting procedure accelerates momentum transfer between cells 
and leads to a collisional contribution to the transport coefficients. 
However, the resulting expressions, while accurate, were only approximate, 
since it was not possible to sum-up in any controlled fashion all the terms  
in the GK relations. Subsequently, Kikuchi {\it et al} \cite{kiku_03} 
used a non-equilibrium approach to derive expressions for the shear viscosity 
which differed slightly from those derived in \cite{ihle_03b}. Furthermore, 
while their approach yielded only two---pure kinetic and 
collision---contributions to the viscosity, the analysis of the GK formalism 
presented in Refs. \cite{ihle_03a} and \cite{ihle_03b} suggested that there 
are additional ``mixed'' contributions. These discrepancies led us to  
re-examine the GK approach.   

In this paper we show that it is possible to resum the time series 
in the GK relation in such a way as to eliminate all dependence on the  
particles' space-fixed cell coordinates. This leads to dramatic 
simplifications and allows the exact evaluation of the collisional 
contribution to the transport coefficients. Furthermore, it is shown 
that there are only pure kinetic and collision contributions to the transport 
coefficients, with no cross terms. 

In the SRD algorithm, the fluid is modeled by particles with continuous 
spatial coordinates ${\bf r}_i(t)$ and velocities ${\bf v}_i(t)$. 
The system is coarse-grained into the cells of a regular lattice with
no restriction on the number of particles in a cell. The evolution of
the system consists of two steps: streaming and collision. In the
streaming step, the coordinate of each particle is incremented by its
displacement during the time step, $\tau$. Collisions are
modeled by a simultaneous stochastic rotation of the relative velocities
(relative to the mean velocity of the particles in a cell)
of {\em every} particle in each cell.
As discussed in Refs. \cite{ihle_01} and \cite{ihle_03a}, a random shift 
of the particle coordinates before the collision step is required to 
ensure Galilean invariance. 
All particles are shifted by the {\em same} random vector with 
components in the interval $[-a/2,a/2]$ before the collision step. 
There is a great deal of freedom in how the 
rotation step is implemented, and any stochastic rotation matrix consistent 
with detailed balance can be used. 
In two dimensions, the stochastic rotation matrix 
is typically 
taken to be a rotation by an angle $\pm\alpha$, with probability $1/2$ 
\cite{male_99}. In three dimensions, two 
collision rules, denoted by Models A and B in Ref.~\cite{tuze_03},  
have been considered. In Model A \cite{male_00a}, one performs rotations by 
an angle $\alpha$ about a randomly chosen axis. In Model B \cite{tuze_03}, 
rotations are performed about one of three orthogonal rotation axes  
of a cartesian coordinate system. At each collision step, one of these three 
axes is chosen at random, and a rotation by an angle $\pm\alpha$ is then 
performed, where the sign is chosen at random. 
	
Explicit GK relations for the transport coefficients of the SRD algorithm 
were derived in Ref. \cite{ihle_03a}.
In particular, it was shown that the shear viscosity, $\nu$, is given by  
\cite{ihle_03b,tuze_03}
\begin{equation}
\label{VIS1}
\nu = {\tau \over N \,k_B T} 
\left.\sum_{n=0}^\infty \right.'
\langle \sigma_{xy}(0) \sigma_{xy}(n \tau)\rangle\ , 
\end{equation}
where 
\begin{eqnarray}
\label{VIS2}
\sigma_{xy}(n\tau)=&& \nonumber \\
&&\!\!\!\!\!\!\!\!\!\!\!\!\!\!\!\!\!\!\!\!\!\!\!\!\!\!\!\!\!\!\!\!-{1\over\tau}\sum_j[v_{jx}(n \tau)\Delta\xi_{jy}(n\tau) +
                           \Delta v_{jx}(n\tau)\Delta\xi^s_{jy}(n\tau)]\ , 
\end{eqnarray}
with $\Delta\bxi_j\left(n\tau\right) = 
\bxi_j\left(\left[n+1\right]\tau\right) - 
\bxi_j\left(n\tau\right)$, $\Delta\bxi^s_j\left(n\tau\right) = $
$\bxi_j\left(\left[n+1\right]\tau\right) - 
\bxi^s_j\left(\left[n+1\right]\tau \right)$, 
and 
$\Delta v_{xj}(n\tau)=v_{xj}\left(\left[n+1\right]\tau\right)-v_{xj}(n\tau)$. 
$\bxi_j(n\tau)$ is the cell coordinate of particle $j$ at time $n\tau$, 
while $\bxi_j^s$ is it's cell coordinate in the (stochastically) shifted 
frame.  
The prime on the sum indicates that the $t=0$ term has the relative weight 
$1/2$. The sum in Eq. (\ref{VIS2}) runs over all $N$ particles of the system. 
Here and in the following we have set the particle mass equal to one.  
\begin{figure}[t]
\begin{center}
\leavevmode
\includegraphics[width=3in,angle=0]{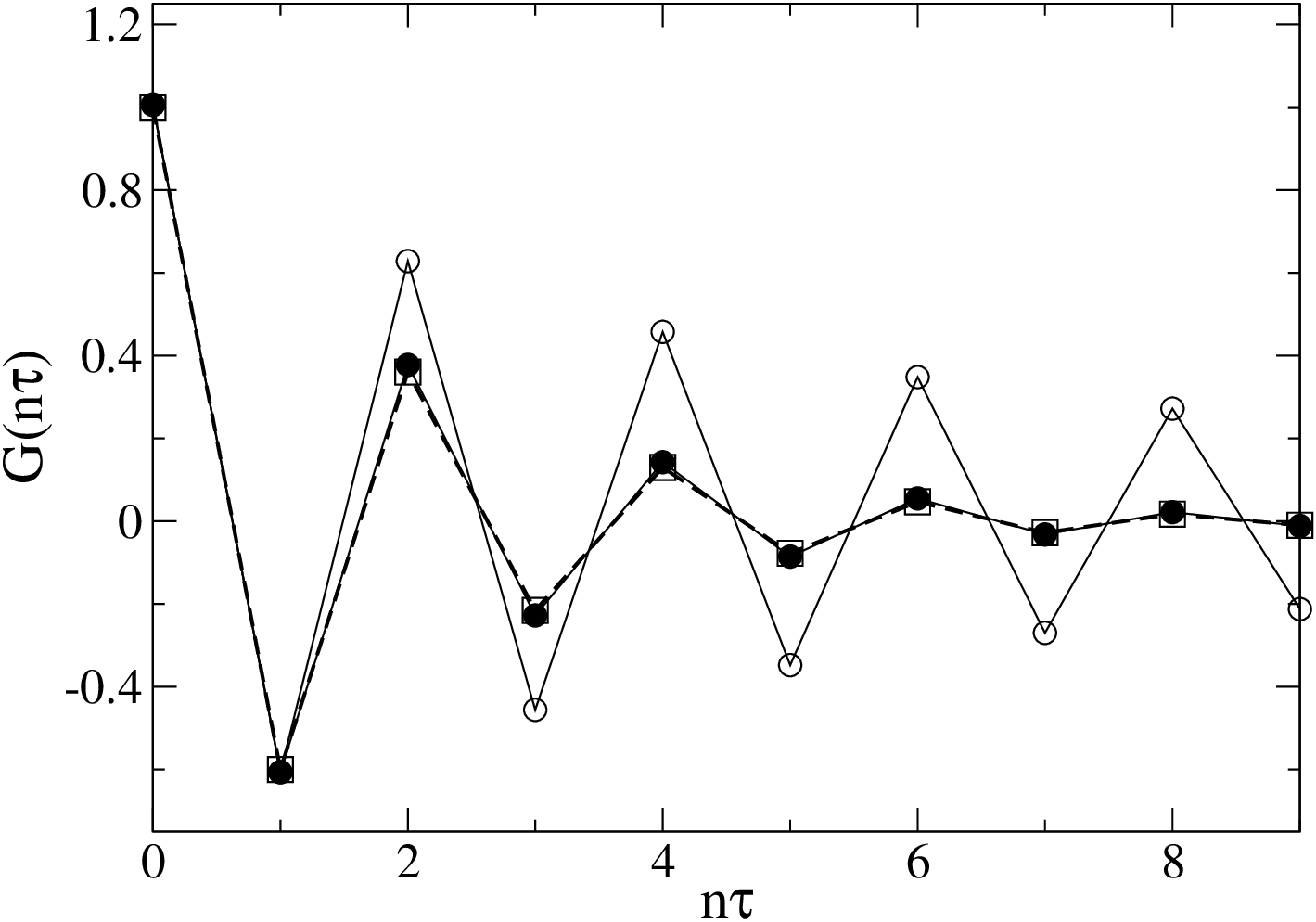}
\caption{$G(n\tau)$
as a function of time step for $\lambda/a=0.01$ ($\circ$)
and $\lambda/a=1.0$ ($\ssquare$). The bullets ($\bullet$) are the result 
$G_C(n\tau)$ given in Eq. (\ref{G_C}).  
Averages were taken 100,000 iterations and 5 different random number seeds. 
Parameters: $L/a=64$, $M=5$, and $\alpha=90^o$.}
\label{series_alpha90}
\end{center}
\end{figure}

The straightforward evaluation of the GK relations presented in Ref. 
\cite{ihle_03b} leads to three contributions to the transport coefficients, 
which were called the kinetic, rotational and mixed terms. For large 
mean free path, $\lambda=\tau \sqrt{k_B T} \gg a$, the assumption of molecular chaos is valid, 
and the kinetic contribution could be determined explicitly. For mean
free paths smaller than the cell size, however, there are finite cell 
size corrections, and it was not possible to sum these contributions 
in a controlled fashion. The origin of the problem was the explicit 
appearance of the cell coordinate $\Delta \xi$ in the stress correlation 
functions.   

In fact, the appearance of $\Delta \xi$ is troubling, since one would 
not expect this to be the case if the cell shifting procedure really does 
restore Galilean invariance. The key to resolving this dilemma is to 
realize that a proper resummation 
of the GK relations removes this dependence. In particular,  
by canceling $\xi$-dependent terms in successive contributions 
to the time series in Eq. (\ref{VIS1})
and using stationarity \cite{tuze_04},   
it can be shown that transport coefficients are given by the same GK 
relations, but with the stress tensor 
$\sigma_{xy}(n\tau)\equiv\bar\sigma^{kin}_{xy}(n\tau)+
\bar\sigma^{rot}_{xy}(n\tau)$, 
with 
\begin{equation}
\label{VIS2Na}
\bar\sigma^{kin}_{xy}(n\tau)=-\sum_j v_{jx}(n\tau)v_{jy}(n\tau),   
\end{equation} 
and 
\begin{equation} 
\label{VIS2Nb}
\bar\sigma^{rot}_{xy}(n\tau) = -\frac{1}{\tau}\sum_j 
B_{jy}(n\tau)v_{jx}(n\tau),
\end{equation} 
where  
$B_{j\beta}(n \tau)=\xi_{j\beta}^s\left(\left[n+1\right]\tau\right)
                    -\xi_{j\beta}^s(n\tau)-\tau v_{j\beta}(n\tau)$.
\begin{figure}[t]
\begin{center}
\leavevmode
\includegraphics[width=3in,angle=0]{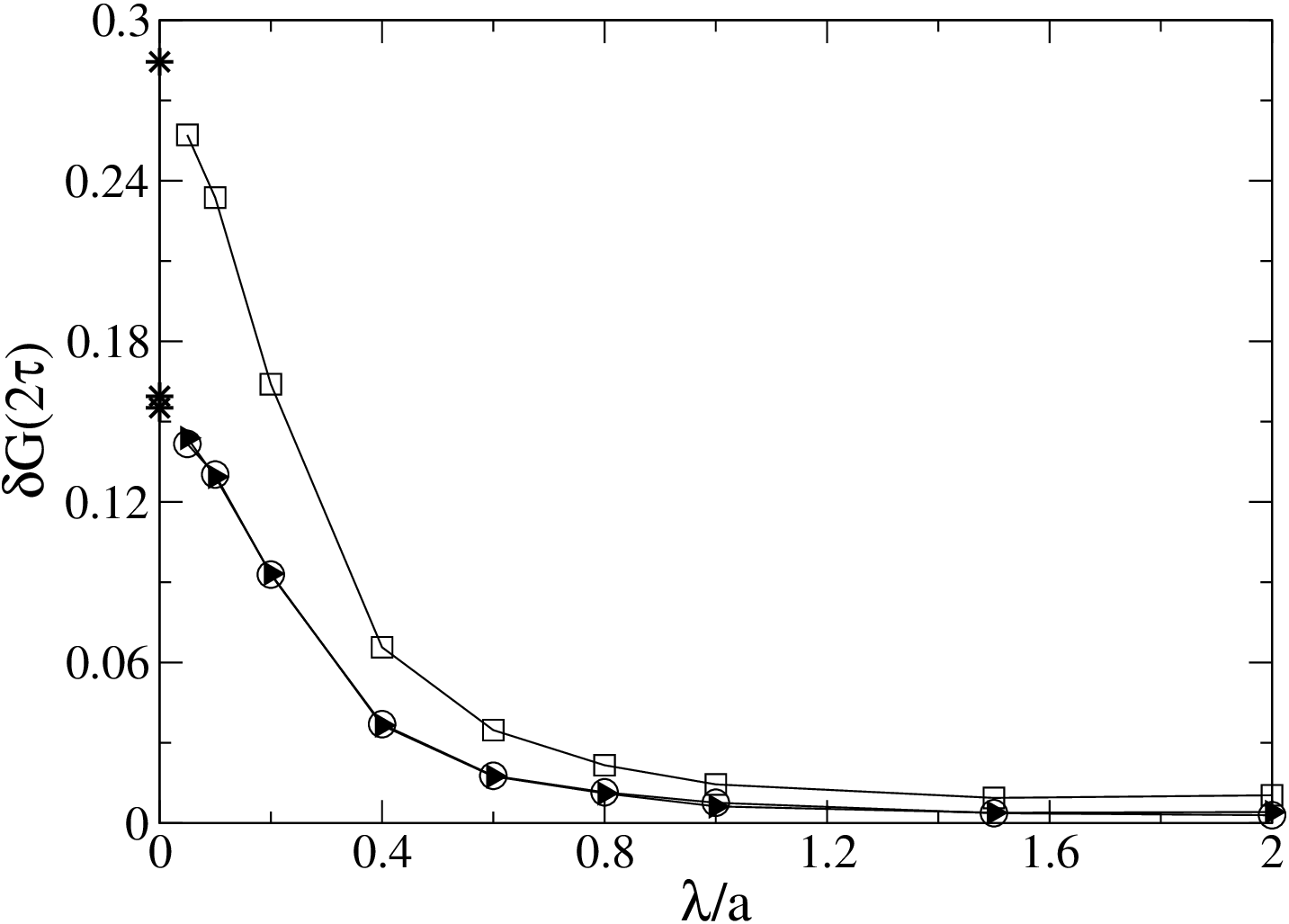}
\caption{$\delta G(2\tau)$ as a function 
of $\lambda/a$ for rotation angles $\alpha=60^\circ$ ($\lcirc$), 
$90^\circ$ ($\ssquare$), and $120^\circ$ ($\blacktriangleright$).   
The asterisks ($*$) are the theoretical values values for $\lambda/a=0$.  
Time averages over 400,000 iterations for 5 different random number seeds 
were used to obtain the data. Parameters: $L/a=64$ and $M=5$.}
\label{fig_relcorr}
\end{center}
\end{figure}
Note that the new stress tensor does not depend on $\xi$, the space-fixed 
cell coordinates of the particles.   
It can be shown \cite{tuze_04}, and has been verified numerically, that  
$\langle B_{i\alpha} \rangle = 0$ and that all correlations of the $B$-fields 
with the particle velocities in the stress correlation functions factorize.   
Furthermore,  
\begin{eqnarray}
\label{VIS8}
\langle B_{i \alpha}(n\tau) B_{j \beta}(m\tau) \rangle= &&\nonumber \\
&&\!\!\!\!\!\!\!\!\!\!\!\!\!\!\!\!\!\!\!\!\!\!\!\!\!\!\!\!\!\!\!\!\!\!\!\!\!\!\!\!\!\!\!\!\!\!\!\!\!\!\! {a^2 \over 12}
\, \delta_{\alpha\beta} (1+\delta_{ij})
\left[2 \delta_{n,m} - \delta_{n,m+1}-\delta_{n,m-1}  \right] , 
\end{eqnarray}
so that the $B$'s are uncorrelated for time lags greater than one time step. 
These relations imply that there are only two---a pure kinetic and a pure 
rotational  ---contributions to the transport coefficients. 
Relation (\ref{VIS8}) is of central importance, because it contains all the 
geometrical features of the grid that contribute to the transport 
coefficients, and is independant of specific collision rules and particle
properties.

Using these results in (\ref{VIS1}), the viscosity can be 
written as $\nu=\nu_{kin}+\nu_{rot}$, with
\begin{equation}
\label{VIS9}
\nu_{kin}  =   {\tau \over N \,k_B T} \left.\sum_{n=0}^\infty \right.'
\sum_{i,j=1}^N \langle v_{xi}(0) v_{yi}(0)v_{xj}(n\tau) v_{yj}(n\tau)\rangle 
\end{equation} 
and
\begin{eqnarray} 
\label{VIS10}
\nu_{rot}  &=&  {\tau \over 2 N \,k_B T}  \sum_{i,j=1}^N 
\left\{
\langle v_{ix}(0) v_{jx}(0) \rangle \langle B_{iy}(0) B_{jy}(0) \rangle \right. \nonumber \\
~~~~~~~~~ &+& \left. 2 \langle v_{ix}(0) v_{jx}(\tau) \rangle \langle B_{iy}(0) B_{jy}(\tau) 
\rangle \right\} \;.
\end{eqnarray}
\begin{figure}[t]
\begin{center}
\leavevmode
\includegraphics[width=3in,angle=0]{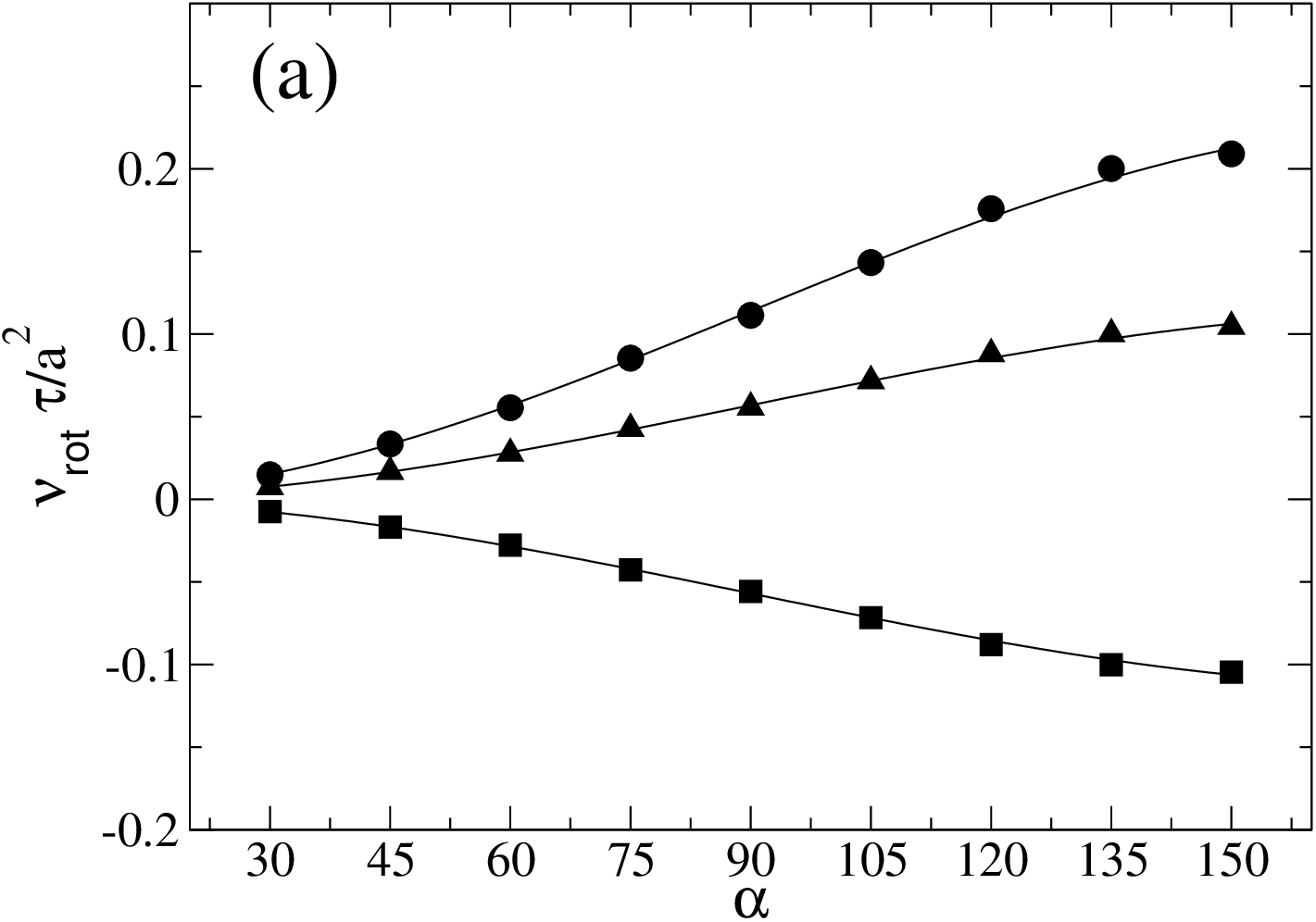}
\end{center}
\vspace{0.2cm}
\begin{center}
\includegraphics[width=3in,angle=0]{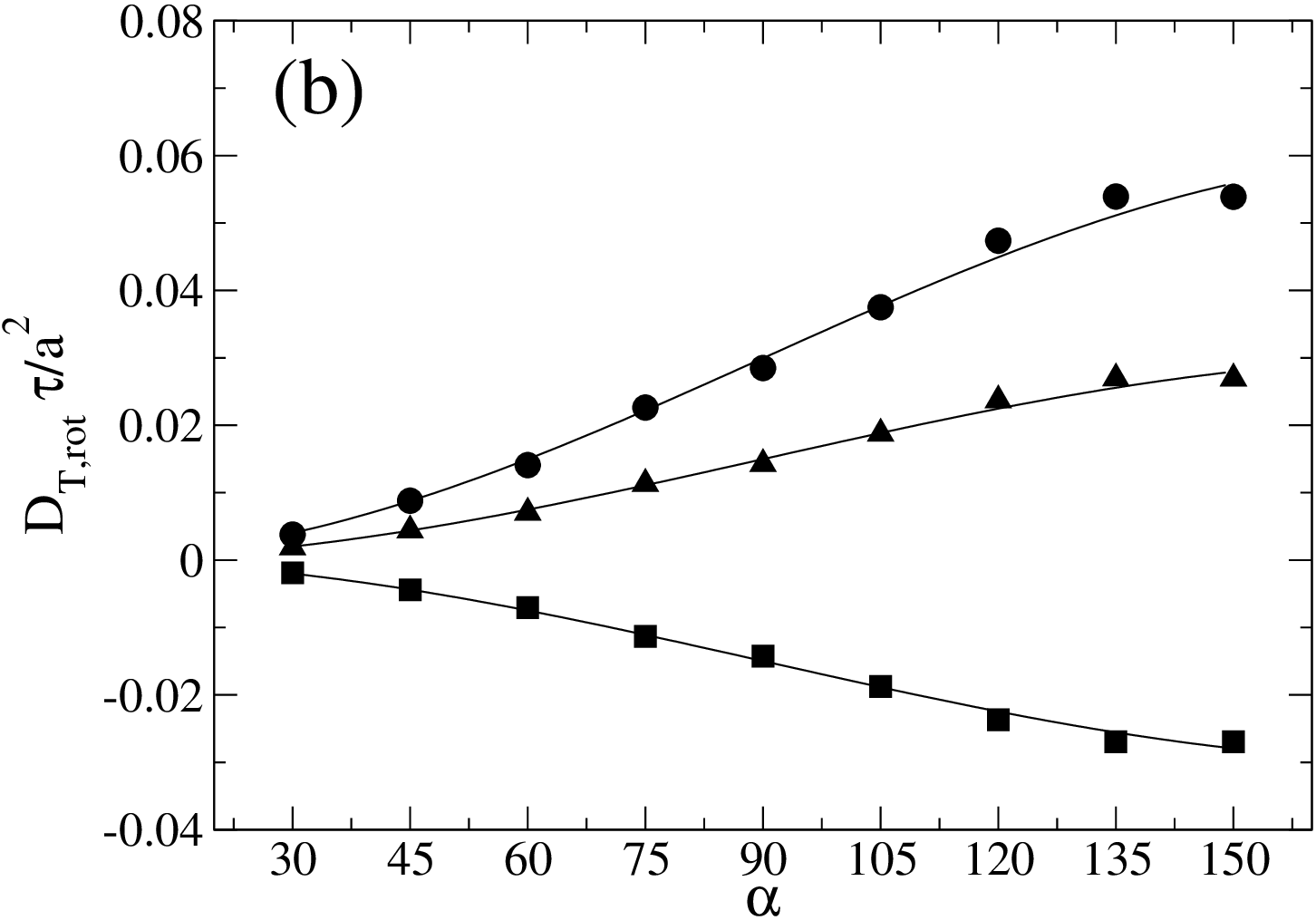}
\caption{Simulation results for the normalized rotational contribution to 
a) the kinematic viscosity, $\nu_{rot}\tau/a^2$, and b) the thermal 
diffusivity, 
$D_{T,rot}\tau/a^2$, as a function of the collision angle $\alpha$.
The bullets ($\bullet$) are the diagonal, the squares ($\blacksquare$) 
the off-diagonal, and the triangles ($\blacktriangle$)  the total contribution to the rotational 
viscosity and thermal diffusivity. The solid lines are the theoretical predictions.  
The data were obtained by time averaging over 360,000 iterations.
Parameters: $L/a=16$, $\lambda/a=0.1$, $M=3$ and $\tau=1$.}
\label{fig_nudt_rot}
\end{center}
\vspace{-0.5cm}
\end{figure}
Assuming molecular chaos, it is straightforward to evaluate the kinetic 
contribution to the shear viscosity. 
If, in addition,  it is assumed that the 
number of particles in any cell is Poisson distributed at each time step, 
with an average number $M$ of particles per cell, and average over the 
number of particles in a cell~\cite{kiku_03}, one finds  
\begin{eqnarray} 
\label{nu_kin} 
\nu_{kin}^{2D} &=& k_BT\tau \left.\sum_{n=0}^\infty \right.'
\,G_C(n\tau) \nonumber \\
&=&
\frac{k_BT\tau}{2}\left[\frac{M}{(M-1+e^{-M})\sin^2(\alpha)}
             -1\right],  
\end{eqnarray} 
in two dimensions, where 
\begin{eqnarray}
\label{G_C} 
G_C(n\tau)&\equiv&  
\langle \bar\sigma^{kin}_{xy}(0) \bar\sigma^{kin}_{xy}(n \tau)\rangle_C / 
N(k_BT)^2  \nonumber \\ 
&=& [1-2\sin^2(\alpha)(M-1+e^{-M})/M]^n.
\end{eqnarray}
The index $C$ indicates that molecular chaos was assumed when performing 
the averages. 
The corresponding result for Model A in three dimensions is 
\begin{eqnarray}
\label{nu_kin3}
\nu_{kin}^{3D} =&& \\
&&\!\!\!\!\!\!\!\!\!\!\!\!\!\!\!\!\!\!\!\!\! \!\!\!\!\!\! \frac{k_BT\tau}{2}
\left[\frac{5M}{(M-1+e^{-M})[2-\cos(\alpha)-\cos(2\alpha)]}
             -1\right] \nonumber  
\end{eqnarray}
(see also Refs. \cite{tuze_03,tuze_04}). 

Eqs. (\ref{nu_kin}) and (\ref{nu_kin3}) are the same results one would  
obtain in the Chapman-Enskog approximation \cite{male_99}. 
For small mean free path, however, there are significant contributions 
to $\nu_{kin}$ which are neglected in this approximation. They arise 
from correlations between particles which are in the same (shifted) cell 
at more than one time step.
Fig. \ref{series_alpha90} contains a plot of the $G(n\tau)\equiv 
\langle \bar\sigma^{kin}_{xy}(0) \bar\sigma^{kin}_{xy}(n\tau)\rangle/N(k_BT)^2$,
in two dimensions for $\alpha=90^\circ$.   
The bullets are the result $G_C(n\tau)$ given in (\ref{G_C}), 
and the open squares are simulation data for $\lambda/a=1$; the  
agreement shows that for this value of the mean free path, 
Eq. (\ref{nu_kin}) provides an excellent approximation for $\nu_{kin}$. 
On the other hand, data obtained for $\lambda/a=0.01$ ($\circ$) 
exhibit much larger 
correlations for $n\tau\ge2$. Fig. \ref{fig_relcorr} contains a plot of 
the relative difference, $\delta G(2\tau)=G(2\tau)-G_C(2\tau)$ 
as a function of the mean free path.
While it is rather difficult to evaluate these corrections 
analytically for general $\lambda$, we have calculated $\delta G(2\tau)$ 
in the $\lambda\to0$ limit. The results of this calculation, which are 
shown in Fig. \ref{fig_relcorr}, 
are in excellent agreement with the numerical results.  

There are corrections of this type at small $\lambda/a$ for all the transport 
coefficients, and 
it is important to note that they provide a particularly large contribution 
to the bare self-diffusion coefficient \cite{ripo_04,ihle_03b,tuze_04}. 
The effect of these correlations on the value of the viscosity 
are less significant and only visible at intermediate 
mean free path, since they vanish for large $\lambda$ and are small 
compared to the dominant collisional contribution for $\lambda\ll a$.
For $\lambda/a=0.4$, the correlations at $n=2$ make an  
additional contribution of approximately $12\%$ to the total viscosity.

The rotational contribution to the viscosity is easy to evaluate, 
since, as can be seen from Eq. (\ref{VIS10}), only stress correlation functions 
at equal time and for a time lag of one time step are required. 
Another simplifying feature is that because of momentum conservation, 
the diagonal (from $i=j$) and off-diagonal (from $i\ne j$) contributions 
to $\nu_{rot}$ in (\ref{VIS10}) obey the relation 
$\nu_{rot}^{diagonal}=-2 \nu_{rot}^{o\!f\!f\!-diagonal}$. Using this result 
and relation (\ref{VIS8}), and averaging over the number of particles 
in a cell, one obtains \cite{tuze_04}   
\begin{equation}
\label{VIS11} 
\nu_{rot}=\frac{a^2}{6 d \tau} \left(\frac{M-1+e^{-M}}{M}\right)
[1-\cos(\alpha)] \label{nu_rot_2df} \;,
\end{equation}
for all the collision models we considered (the standard model in $d=2$ 
and both models A and B in $3d$ \cite{tuze_03}).
Eq. (\ref{VIS11}) agrees with the result of Kikuchi {\it et al} 
\cite{kiku_03} obtained using a different non-equilibrium 
approach in shear flow, but deviates slightly for small $M$ from the 
result given in Refs. \cite{ihle_03b} and \cite{tuze_03}. 
Result (\ref{VIS11}) is compared with simulation data for the rotational 
contributation to the viscosity in Fig. \ref{fig_nudt_rot}a. 

The GK relation for the thermal diffusivity, $D_T$, derived in 
Ref. \cite{ihle_03a,ihle_03b} can be resummed
in a similar fashion. In particular, 
it can then be shown that  $D_T=D_{T,kin}+D_{T,rot}$.
$D_{T,kin}$ was calculated in $2d$ in Ref. \cite{ihle_03b} and in $3d$ in 
\cite{tuze_03} neglecting fluctuations in  
the number of particles in a cell. As for the viscosity, it is straightforard 
to include  
particle number fluctuations by averaging the contributions to the 
heat-flux correlation functions over the number of particles in a cell;
the resulting expression will be given elsewhere \cite{tuze_04}. 
The relation 
$D_{T,rot}^{diagonal}=-2 D_{T,rot}^{o\!f\!f\!-diagonal}$, which follows from 
energy conservation, can be used to show that the rotational contribution to 
the thermal diffusivity is 
\begin{widetext}
\begin{equation}
\label{DT_rot} 
D_{T,rot}=\frac{a^2}{3 (d+2) \tau} \frac{1}{M} 
\left[1-e^{-M}\left(1+\int_0^M \frac{e^x-1}{x}dx \right) \right] 
[1-\cos(\alpha)] \approx 
\frac{a^2}{3 (d+2) \tau} \frac{1}{M} \left(1 - \frac{1}{M}\right) 
[1-\cos(\alpha)]      
\end{equation} 
\end{widetext}
to leading order for large $M$. Note that in contrast to the viscosity, 
the rotational contribution to the thermal diffusivity is $O(1/M)$, so that
the corrections to $D_T$ at small $\lambda/a$ arising from 
correlated collisons are more important than for the viscosity. 
Simulation results for $D_{T,rot}$ 
are compared with Eq. (\ref{DT_rot}) in Fig. \ref{fig_nudt_rot}b.

It is now clear that the random shift procedure introduced in Refs. 
\cite{ihle_01} and \cite{ihle_03a} not only restores Galilean invariance, 
but also enables an exact evaluation of the collisional contribution to 
the transport coefficients and clarifies several aspects of the underlying 
algorithm. In addition, the current approach justifies in detail several 
assumptions used in the non-equilibrium calculation of Kikuchi {\it et al} 
\cite{kiku_03} which led them to the same, correct results for the shear 
viscosity derived here using GK relations.
An advantage of the current approach is that it can be used to analyze the
transport coefficients of the longitudinal modes, namely the bulk
viscosity and thermal diffusivity, which are hard to calculate
in a non-equilibrium approach \cite{pool_03}.
It can also be used to show that the bulk viscosity is 
equal to zero \cite{ihle_03b,tuze_04}.

We thank J. Yeomans for helpful discussions which initiated this 
reexamination of the GK approach. We thank her and C.M. Pooley
for making their unpublished notes available to us. Support from the 
National Science Foundation under grant Nos. DMR-0083219 and DMR-0328468, 
the donors of The Petroleum Research Fund, administered by the ACS,  
is greatfully acknowledged.

\end{document}